\documentclass[preprint,prd,nofootinbib]{revtex4}
\usepackage[dvips]{graphicx}
\usepackage{subfigure}
\usepackage{latexsym}
\newcommand{\nc}{\newcommand}
\nc{\beq}{\begin{equation}}
\nc{\eeq}{\end{equation}}
\nc{\beqa}{\begin{eqnarray}}
\nc{\eeqa}{\end{eqnarray}}
\nc{\bert}{\raise-.55mm\hbox{\large$\Box$}}  %D'Alambertian

  \def\lsim{\mathrel{\rlap{\lower4pt\hbox{\hskip1pt$\sim$}}
\raise1pt\hbox{$<$}}}

\def\gsim{\mathrel{\rlap{\lower4pt\hbox{\hskip1pt$\sim$}}
    \raise1pt\hbox{$>$}}}

\newwrite\ffile\global\newcount\figno \global\figno=1

\def\writedef#1{}
\def\figin{\epsfcheck\figin}\def\figins{\epsfcheck\figins}

\def\epsfcheck{\ifx\epsfbox\UnDeFiNeD \message{(NO epsf.tex, FIGURES
WILL BE IGNORED)}
\gdef\figin##1{\vskip2in}\gdef\figins##1{\hskip.5in}% blank space
instead \else\message{(FIGURES WILL BE INCLUDED)}%
\gdef\figin##1{##1}\gdef\figins##1{##1}\fi} \def\figinsert{}
\def\ifig#1#2#3{\xdef#1{fig.~\the\figno} \writedef{#1\leftbracket
fig.\noexpand~\the\figno}%
\figinsert\figin{\centerline{#3}}\medskip\centerline{\vbox{\baselineskip12pt
\advance\hsize by -1truein\center\footnotesize{  Fig.~\the\figno.} #2}}
\bigskip\endinsert\global\advance\figno by1}
\def\endinsert{}
\begin{document}

\title{~\\ Gradient Instability for $w<-1$}

\author{Stephen~D.H.~Hsu\footnote{Permanent address: Institute of
Theoretical Science and Department of Physics, University of Oregon,
Eugene, OR 97403.  E-mail: hsu@duende.uoregon.edu},
Alejandro~Jenkins\footnote{jenkins@theory.caltech.edu}, and
Mark~B.~Wise\footnote{wise@theory.caltech.edu}}

\affiliation{California Institute of Technology, Pasadena, CA 91125
\bigskip \bigskip \bigskip \bigskip}

\preprint{OITS-750} \preprint{CALT-68-2493}

\begin{abstract} \bigskip

We show that in single scalar field models of the dark energy with
equations of state satisfying $w \equiv p / \rho < -1$, the effective
Lagrangian for fluctuations about the homogeneous background has a
wrong sign spatial kinetic term. In most cases, spatial gradients are
ruled out by microwave background observations.  The sign of $w+1$ is
not connected to the sign of the time derivative kinetic term in the
effective Lagrangian.

\end{abstract}

\maketitle

%%%%%%%%%%%%%%%%%%%%%%%%%%%%%%%%%%%%%%%%%%%%%%%%%%%%%%%%%%%%%%%%%
%%%
%%%                     INTRODUCTION
%%%
%%%%%%%%%%%%%%%%%%%%%%%%%%%%%%%%%%%%%%%%%%%%%%%%%%%%%%%%%%%%%%%%%

\newpage

Matter whose equation of state satisfies $w \equiv p / \rho < -1$
violates a number of conditions, including the weak energy condition,
generally assumed to apply to any reasonable model of physics
\cite{SMC}. However, the  observational data do not exclude the
possibility that the dark energy has $w < -1$
\cite{Hannestad:2002ur,Melchiorri:2002ux}.  Current results
\cite{Knop2003iy} indicate  $-1.67 < w < -0.61$ at 95\% confidence
level. The possibility of $w < -1$ has been explored by numerous
authors (see, for instance,
\cite{Caldwell:1999ew}--\cite{Holdom:2004yx}). These models often
contain a field with an unusual kinetic term, which is referred to as
a phantom or ghost field.  In this letter we show that for $w < -1$,
single scalar field models of the dark energy generally have a wrong
sign gradient kinetic term for fluctuations about the homogeneous
background. This result is not dependent on general relativistic
effects and survives in the flat spacetime limit.  Spatial
inhomogeneities of the dark energy are tightly constrained by
observations of the cosmic microwave background.

\begin{figure}[t]
\begin{center}
\includegraphics[bb=135 510 270
630,clip,width=.3\textwidth]{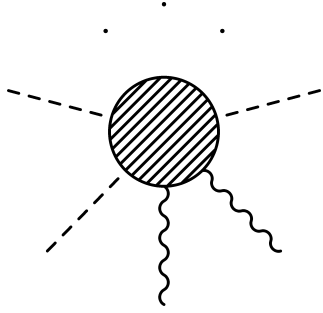}
\caption{The effective couplings of two gravitons to several quanta of
the scalar field.  The shaded region represents interactions involving
only scalars.}
\label{GhostGravitonBlob}
\end{center}
\end{figure}

In our analysis we will assume a time-dependent but spatially
homogeneous scalar background.  Models with such backgrounds are
treated in \cite{ghost1}, whose notation we will adopt.  We will show
that for $w<-1$ spatial instabilities inevitably arise.  Consider the
low-energy effective theory of a scalar field coupled to gravity: \beq
\label{nonminimalcoupling} S = \int d^4x \sqrt{-g}  \left[ M_{Pl}^2
R + P + U \, R + V \, R^{\mu\nu} (\partial_\mu \phi)(\partial_\nu
\phi) + ~\cdots ~\right]~, \eeq where $P, U$ and $V$ are functions of
the scalar field $\phi$ and its derivatives.  (Because of the
anti-symmetry of $R^{\mu\nu\rho\sigma}$ in its first two and also in
its last two indices, no non-vanishing invariant can be formed from it
using first derivatives of $\phi$.)  Naively we might expect that the
higher-dimensional couplings of $\phi$ to the Ricci tensor would be
suppressed by powers of the Planck mass $M_{Pl}$, making them
irrelevant for cosmology after the Planck epoch.  However, such terms
are generated by graphs such as that in Figure
\ref{GhostGravitonBlob}.  Writing the metric as $g^{\mu\nu} =
\eta^{\mu\nu} + h^{\mu\nu} / M_{Pl}$, we see that scalar-graviton
interactions in Feynman diagrams are suppressed by the Planck mass,
but when these interactions are reassembled into the Ricci tensor that
suppression is absent.  That is, the higher-dimensional terms in
Eq. (\ref{nonminimalcoupling}) will appear suppressed  only by powers
of the characteristic energy scale of the scalar field, $M$, which may
be much smaller than $M_{Pl}$.

We neglect terms in the action (\ref{nonminimalcoupling}) which
involve higher powers of the Ricci tensor.  The terms we consider are
ones that can generate contributions to the stress-energy tensor
$T_{\mu\nu}$ which are not suppressed by powers of $M_{Pl}$.  Since
$T_{\mu\nu}$ is obtained by varying the action with respect to the
metric, terms with more than one power of $R^{\mu\nu\rho\sigma}$ yield
contributions which are themselves proportional to the Ricci tensor
and therefore vanish in the flat spacetime limit.

Assuming a spatially homogeneous background, only the time-derivatives
of $\phi$ will be non-vanishing in Eq. (\ref{nonminimalcoupling}).  It
may be shown that in the limit $M_{Pl} \rightarrow \infty$, the term
$R^{\mu\nu} \, (\partial_\mu \phi)(\partial_\nu \phi)V$ contributes a
term to the stress-energy tensor which can be reproduced by an
appropriate change in the function $U$.  Therefore we may restrict
ourselves to $V=0$ and consider the most general $U$ in order to
analyze the flat-space behavior of Eq. (\ref{nonminimalcoupling}).

It is always possible to perform a rescaling of the metric in
Eq. (\ref{nonminimalcoupling}) $g_{\mu\nu} \rightarrow e^{2w}
g_{\mu\nu}$, with $w = \log[1 + U / M_{Pl}^2]$, so that  the $U$ term
in Eq. (\ref{nonminimalcoupling}) disappears, being absorbed into a
redefinition of the $P$ action for the ghost scalar field.  (See, for
example, Ch. 3, Sec. 7 of \cite{Polchinski}.)  The action resulting
from this rescaling, up to terms suppressed by powers of $1 / M_{Pl}$,
is then \beq S = \int d^4 x \sqrt{-g} \left[ M_{Pl}^2 R + P
\right]~. \label{Einsteinframe} \eeq

The most general Lorentz invariant scalar Lagrangian without higher
derivative terms (which we will consider later) is \beq
\label{model} {\cal L} = P(X,\phi)~~, \label{PXX} \eeq where $X =
g^{\mu\nu}\partial_\mu \phi \, \partial_\nu \phi$.  (A potential term
$V$ would be the component of $P(X, \phi)$ that is independent of
$X$.)  Henceforth, $P'(X,\phi)$ will denote differentiation with
respect to $X$.   Since the scalar field $\phi$ is minimally coupled
to gravity in Eq.  (\ref{Einsteinframe}), the stress-energy tensor is
\beq
\label{T} T_{\mu \nu} = - {\cal L} g_{\mu \nu} + 2 P'(X,\phi)
\partial_\mu \phi \partial_\nu \phi~~,\eeq and \beq
\label{WPP}
w = { {P(X,\phi)} \over {T_{00}}} = {{P(X,\phi)} \over {-P(X,\phi) +
2{\dot{\phi}}^2P'(X,\phi)}} = -1 + {{2 {\dot{\phi}}^2P'(X,\phi)} \over
{T_{00}}}~~.  \eeq  For $\phi$ to account for the dark energy, we must
have $T_{00} > 0$.  Then, $w < -1$ requires that $P'(X,\phi) < 0$. Let
$\phi_0=\phi_0(t)$ be a solution to the equations of motion, and
define $X_0 \equiv \dot\phi_0^2$.  Then consider the fluctuations
about this solution: $\phi = \phi_0 + \pi (x,t)$. When expanded in
$\pi$, the effective Lagrangian will have the form \beq
\label{spatial} {\cal L} = \left[P'(X_0,\phi_0) + 
2 \dot\phi_0^2 P''(X_0,\phi_0)\right] \dot\pi^2 - P'(X_0,\phi_0) \vert
\nabla \pi \vert^2 + \cdots~, \eeq which implies that for
$P'(X_0,\phi_0) < 0$ there will exist field configurations with
non-zero spatial gradients that have lower energy than the homogeneous
configuration.\footnote{Here we mean energy associated with the
Hamiltonian constructed from the Lagrangian for fluctuations about the
background field configuration.}

It is clear from Eq. (\ref{spatial}) that there is no direct
connection between the sign of $w+1$ and that of the $\dot \pi^2$ term
in the effective Lagrangian, as long as $\dot\phi_0^2 P''(X_0,\phi_0)$
is not negligible with respected to $P'(X_0,\phi_0)$.  However, in the
usual quintessence models of the dark energy (see \cite{quintessence}
and references therein), it is generally the case that the terms in
$P$ which are higher order in $X$ can be neglected with respect to the
leading term (i.e., $X/M^4 \ll 1$).  In that case the $\dot \pi^2$ and
$-(\nabla \pi)^2$ terms in the effective Lagrangian would both have a
negative coefficient for $w < -1$.

If $P'(X_0,\phi_0)$ is negative, a finite expectation value for the
gradients may be obtained if there are appropriate higher powers of
$(\nabla \pi)^2$ in the effective Lagrangian, but this is problematic
because it gives rise to a spatially inhomogeneous ground state for
the dark energy and would lead to inhomogeneities far larger than the
limit of $10^{-5}$ imposed by observations of the cosmic microwave
background.\footnote{A condensate of gradients with a preferred
magnitude, determined by the higher order terms that stabilize
Eq. (\ref{spatial}), will spontaneously break the $O(3)$ rotational
symmetry down to $O(2)$.  The homotopy group $\pi_2[O(3)/O(2)]$ is
non-trivial, which leads to the formation of global monopole
(hedgehog) configurations.} While a potential term such as $m^2
\phi^2$ tends to confine the gradients to regions of size $1/m$, in
most models of the dark energy $V'' (\phi)$ must be small enough that
these regions are of cosmological size.

In the $w<-1$ case, it is possible, by adding higher derivative terms
to the Lagrangian, to avoid having finite spatial gradients lower the
energy of the field.  Consider, for example: \beq
\label{higherd}
{\cal L} = P(X,\phi) + S(X,\phi)(\bert \phi)^2 \eeq in which case  \beq
\label{higherdrho}
T_{00} = -{\cal L} + 2[P'(X,\phi)\dot{\phi}^2 + S'(X,\phi)
\dot{\phi}^2 (\partial^2 \phi)^2 + 2S(X,\phi)\ddot{\phi}(\partial^2
\phi) -\partial_0(\dot\phi S \partial^2\phi)]~. \eeq  Setting the
spatial gradients of $\phi$ to zero, we have that \beq \dot{\phi}^2
C_{grad} -2 \partial_0(S\ddot\phi)\dot\phi = {(w+1) \over 2} \,
T_{00}~, \eeq where $C_{grad}$ is the coefficient of $-(\nabla \pi)^2$
in the $\pi$ Lagrangian.  If $\partial_0(S\ddot\phi)\dot\phi>0$, then
a model may have both $C_{grad}>0$ and $w<-1$.  But for $w$
significantly less than $-1$, this also requires $\ddot{\phi}^2$ to be
at least of order $M^2\dot\phi^2$, unless $S(X,\phi)$ is made
unnaturally large.  It is not clear how to treat these higher
derivative terms self-consistently beyond perturbation theory, so the
dynamics of such models cannot be analyzed in a straightforward
manner.  The models we consider below have higher powers of first
derivatives, but they satisfy the condition that $\ddot \phi^2 \ll
(\dot \phi M)^2$.

Our analysis shows that $w<-1$ scalar models typically require a wrong
sign $(\nabla \pi)^2$ term in the effective Lagrangian.  Previous
analyses of ghost models \cite{SMC,MCG} have focused on the problems
associated with negative energy, particularly with a kinetic term
${\cal L} = -(\partial_{\mu} \phi)^2$ that has the wrong sign for {\it
both} the time- and space-derivatives.  The classical equations of
motion for such models do not exhibit growing modes of non-zero
spatial gradients, although the energy of the field is unbounded from
below.  Models with $w<-1$ that do not have a wrong sign
time-derivative kinetic term in the effective Lagrangian can result
from a Lorentz invariant action, as we demonstrate below.   However,
both Lorentz invariance and time translation invariance are
spontaneously broken by a time-dependent condensate.

In \cite{ghost1} a model with ${\cal L} = P(X)$  was proposed in which
a ghost field has a time-dependent condensate (from now on we take the
Lagrangian to be a function of X only, and therefore invariant under
the shift $\phi \rightarrow \phi + c$).  We use units in which the
dimensional scale $M$ of the model is unity ($M \sim 10^{-3}$ eV if
the ghost comprises the dark energy).  The flat spacetime equation
motion is \beq \label{EM} \partial_\mu \left[ P^\prime(X) \partial^\mu
\phi \right] = 0~~. \eeq Homogeneous solutions of the equations of
motion with $\dot{\phi}^2 = c^2$ were considered in \cite{ghost1}.  In
general, the existence of a $\dot{\phi}$ condensate allows for exotic
equations of state, including $w<-1$.  In what follows we let \beq
\label{Pmodel} P(X) = -1 + 4\, (X-1)^2 + 3\,(X-1)^3~, \eeq which leads
to $w < -1$ with $T_{00} > 0$, for X in a left neighborhood of
1.\footnote{Notice that the model in Eq. (\ref{Pmodel}) has a positive
  leading-order kinetic term (i.e., the term linear in X).  It is
  possible to construct models with the desired properties in which
  this sign is either positive or negative.}

The energy density is given by \beq
\label{H}
T_{00} = {\cal H} = {{\partial {\cal L}} \over {\partial \dot{\phi}}}
\dot{\phi} - {\cal L} = 2 \dot{\phi}^2 P^\prime(X) - {\cal L}~~~, \eeq
which is not necessarily minimized by a particular ghost condensate
$\phi = ct$, although it is a solution to the  flat spacetime
equations of motion for any value of $c$.  This is possible  because
there is a conserved charge associated with the shift symmetry, \beq Q
= \int d^3x~ P'(X) \dot{\phi}~~, \eeq so configurations which do not
extremize $T_{00}$ can still be stable.  In fact, the Lagrangian
describing small fluctuations has the correct sign of $\dot{\pi}^2$ if
$P'(X) + 2XP''(X) > 0$. This condition is satisfied in a neighborhood
of $X=1$ by (\ref{Pmodel}) given above. For $c^2 < 1$ there is then a
local instability to the formation of gradients, as required by our
earlier results.

Ghost models of the dark energy which approach $w=-1$ asymptotically
make potentially interesting predictions for the evolution of the
equation of state for the dark energy.  In a FRW universe, the
equation of motion for the ghost field is \beq \label{EMa}
\partial_\mu \left[ a^3(t) \, P^\prime(X) \partial^\mu \phi \right] =
0~~, \label{FRWEoM}\eeq where $a(t)$ is the FRW scale factor.  If
there is a value $c_*^2 = \dot\phi^2 = X$ such that $P'(c_*^2)=0$,
then Eq. (\ref{WPP}) implies that $w=-1$ when $X=c_*^2$.  The model
described by Eq. (\ref{Pmodel}) has $c_*^2=1$, and if we apply
Eq. (\ref{FRWEoM}) to it, we see that if we start from $X=c_i^2$ with
$c_i$ close to $c_*$, then we are driven asymptotically towards
$X=c_*^2$ and $w=-1$.

In the model described by Eq. (\ref{Pmodel}), we may be driven towards
$w=-1$ either from above or from below, depending on whether we chose
to start from $c_i^2>1$ or from $c_i^2<1$.  We have argued that $w<-1$
is problematic because of spatial gradient instabilities, so that the
case in which we are driven to $w=-1$ from above is more interesting.

Near the asymptotic value $c_*=1$ we have:  \beq \dot\pi =
\frac{P'(c_i^2)c_i}{2P''(c_*^2)c_*^2}\left(\frac{a_i}{a}\right)^3~~,
\eeq where higher order terms in $a_i/a$ have been neglected.  Thus,
in this regime, \beq \label{w} w = -1 -
\frac{4P''(c_*^2)c_*^3\dot\pi}{P(c_*^2)} = -1 - \frac{2P'(c_i^2)
c_*c_i}{P(c_*^2)}\left(\frac{1+z}{1+z_i}\right)^3~.\eeq  Equation
(\ref{w}) offers a prediction for the $w$ parameter of the dark energy
as a function of the redshift $z$, which could be tested by
cosmological observation \cite{Linder}.

In summary, from Eqs. (\ref{WPP}) and (\ref{spatial}) we find that in
single scalar field models of the dark energy with $w<-1$, the kinetic
term for fluctuations about the homogeneous background has a  wrong
sign gradient term.  On the other hand, there is no direct connection
between the sign of the $\dot\pi^2$ kinetic term in the effective
Hamiltonian and the sign of $w+1$.

\bigskip

%%%%%%%%%%%%%%%%%%%%%%%%%%%%%%%%%%%%%%%%%%%%%%%%%%%%%%%%%%%%%%%%%
%%%
%%%                   ACKNOWLEDGMENTS
%%%
%%%%%%%%%%%%%%%%%%%%%%%%%%%%%%%%%%%%%%%%%%%%%%%%%%%%%%%%%%%%%%%%%
%\section*{}

\noindent The authors would like to thank N. Arkani-Hamed, X. Calmet,
M. Graesser, M. Kamionkowski, and E. Linder for discussions.  This
work was supported in part under DOE contracts DE-FG06-85ER40224 and
DE-FG03-92ER40701.

%%%%%%%%%%%%%%%%%%%%%%%%%%%%%%%%%%%%%%%%%%%%%%%%%%%%%%%%%%%%%%%%%
%%%
%%%                     BIBLIOGRAPHY
%%%
%%%%%%%%%%%%%%%%%%%%%%%%%%%%%%%%%%%%%%%%%%%%%%%%%%%%%%%%%%%%%%%%%

\bigskip

%\newpage
%\vskip .75 in

\end{document}